\documentclass[prb,aps,twocolumn,amsmath,amssymb,floatfix,superscriptaddress]{revtex4}
\usepackage{graphicx}
\usepackage{bm}
\usepackage{amsmath}
\usepackage{amssymb}
\usepackage{latexsym}
\usepackage{epsf,graphics,graphicx}
\usepackage{comment}
\usepackage{enumitem}

\usepackage[colorlinks=true, citecolor=blue, urlcolor=blue ]{hyperref}

\begin{document}

\title{Interplay of dimerization and quasiperiodicity in the superconducting proximity effect of a one-dimensional hybrid ring}

\author{Sourav Karmakar}
\email{karmakarsourav2015@gmail.com}
\affiliation{Physics and Applied Mathematics Unit, Indian Statistical Institute, 203 Barrackpore Trunk Road, Kolkata-700 108, India}

\author{Santanu K. Maiti}
\email{santanu.maiti@isical.ac.in}
\affiliation{Physics and Applied Mathematics Unit, Indian Statistical Institute, 203 Barrackpore Trunk Road, Kolkata-700 108, India}

\begin{abstract}
Recent studies of the superconducting proximity effect in quasicrystalline and topological systems have opened up a new research direction 
for exploring how quasiperiodicity and topology influence proximity induced superconductivity. In this work, we investigate spatial variation 
of the proximity induced pairing amplitude in a hybrid ring composed of a spin singlet superconductor and a normal region described by 
three different lattice models using the self-consistent Bogoliubov-de Gennes formalism.  We first consider the normal region described 
by the diagonal Aubry-Andr\'e-Harper (AAH) model. Increasing the quasiperiodic potential enhances spatial fluctuations in the induced order 
parameter, while progressively suppressing its magnitude in the normal region. Beyond the localization transition, proximity-induced pairing 
is strongly diminished due to the localized nature of the underlying electronic states. The normal region is then modeled by the
Su-Schrieffer-Heeger (SSH) chain to investigate the effect of hopping dimerization. Weak dimerization introduces oscillatory modulations in 
the induced pairing that extend deep into the normal region, whereas strong dimerization confines these oscillations and significantly 
reduces the penetration of superconducting correlations. Finally, we study the combined SSH-AAH model to explore the interplay between 
the dimerized hopping and quasiperiodicity. The results show that the SSH dimerization determines the oscillatory behavior and penetration 
of the induced pairing, while the AAH potential enhances spatial inhomogeneity and further reduces its magnitude. Together, these two 
effects provide a versatile means of controlling proximity-induced superconductivity in quasiperiodic hybrid systems.
\end{abstract}

\maketitle

\section{Introduction}

The superconducting proximity effect in hybrid structures comprising superconducting and normal-conducting regions has long been 
recognized as a remarkable consequence of quantum coherence in mesoscopic systems. When a superconductor~\cite{sc1} is placed in contact 
with a non-superconducting material, Cooper pairs can penetrate through the interface~\cite{sc2,sc3,sc4,sc5} and induce pairing correlations 
in the adjacent material. The extent of this induced superconductivity depends strongly on whether its electronic states are extended, 
critical, or localized. If the underlying wave functions are extended, superconducting correlations can propagate deep into the normal 
region. In contrast, localized states tend to inhibit this penetration. As a result, the proximity effect provides a sensitive indicator 
of the electronic properties of the non-superconducting systems~\cite{sc6,sc7,sc8,sc9} and offers a useful route for exploring 
unconventional electronic phases.

The nature of electronic states is closely related to the ordering of the underlying lattices. In periodic crystals, translational symmetry
leads to Bloch states that extend throughout the system and support metallic transport. However, in random disordered one-dimensional (1D) 
systems, quantum interference causes Anderson localization~\cite{sc10}, resulting in insulating behavior. In this context, the quasicrystals 
lie between these two paradigms~\cite{sc11, sc12}. Although they possess a long-range order, they lack translational symmetry. Consequently,
their electronic states are generally neither fully extended nor exponentially localized. Instead, they often exhibit critical behavior
characterized by self-similar and multifractal patterns~\cite{sc13,sc14,sc15,sc16,sc17}. This unique nature of quasiperiodic systems 
accounts for a diverse range of physical phenomena that are absent in both perfectly ordered crystals and randomly disordered materials.

A paradigmatic model capturing these features is the Aubry-Andr\'{e}-Harper (AAH) lattice model~\cite{sc18, sc19}, in which a
quasiperiodic modulation of the on-site potential with irrational period drives a sharp transition at a finite critical strength in the
thermodynamic limit. Below this threshold all eigenstates are extended, above it they are all exponentially localized, and precisely at 
the critical modulation strength they are self-similar and multifractal. This transition arises from an exact self-duality of the model 
and is therefore fundamentally distinct from Anderson localization, where arbitrarily weak disorder localizes states in one-dimension. 
The tunability of the AAH model for different electronic states has motivated extensive generalizations, including models with mobility 
edges and extended critical phases~\cite{sc20,sc21,sc22,skm1,skm2,skm3}. The different nature of electronic states of the AAH model 
motivates researchers to investigate how the induced pair correlation is affected by these states in a superconducting-quasiperiodic system.

Another well-known model is the Su-Schrieffer-Heeger (SSH) chain~\cite{sc23,sc24,skm4,skm5}, a one-dimensional dimerized lattice where 
the hopping amplitudes take alternating values of $t_A$ and $t_B$. The SSH model is the prototypical example of a one-dimensional
symmetry-protected topological (SPT) insulator~\cite{sc25}. Its topological phase, realized when $t_A < t_B$, hosts zero-energy edge 
modes at both ends of an open chain~\cite{sc26}. The interplay between the topological structure of the SSH chain and superconducting
correlations has attracted considerable interest because the combination of nontrivial band topology and pairing can give rise to 
topological superconducting phases~\cite{sc27,sc28,sc29,sc30,sc31,sc32,sc33}.

In recent years, the interplay between quasiperiodicity and superconductivity has attracted considerable attention from both theorists 
and experimentalists~\cite{sc34,Sardinero2026,sc35,sc36,sc37,sc38,sc39,sc40,sc41,sc42}. For 1D Fibonacci quasicrystals, in 
which hopping amplitudes follow a Fibonacci substitution sequence with two distinct values, investigations have addressed the proximity
effect~\cite{sc8,sc9}, Josephson transport ~\cite{sc34,Sardinero2026}, and topological superconducting phases~\cite{sc36,sc37,sc41,sc42}. 
Recent studies have shown an enhancement of intrinsic superconductivity in quasiperiodic systems ~\cite{sc36,sc37,sc43}, pointing to a
deep connection between quasiperiodicity and superconducting pairing analogous to the multifractal enhancement of the critical temperature 
in disordered systems. Beyond 1D systems, superconductivity has been studied extensively in two-dimensional quasiperiodic lattices based 
on Penrose and Ammann-Beenker tilings~\cite{sc39,sc40}, where the superconducting gap and local pairing amplitude exhibit a self-similar 
and highly non-uniform spatial pattern governed by the underlying quasiperiodic geometry. An important experimental work was the 
observation of superconductivity in an Al-Zn-Mg quasicrystal~\cite{sc42}, and more recently superconducting behavior was also noticed 
in a van der Waals layered dodecagonal quasicrystal~\cite{sc45}, establishing that quasiperiodic order can support genuine superconducting
phases.

The superconducting proximity effect in quasiperiodic and disordered systems studied by Rai~\cite{sc8,sc9} showed that spatial fluctuations 
in the proximity-induced order parameter increase with decreasing hopping ratio, i.e., with increasing quasiperiodic modulations and their
results establish a direct connection between the localization properties of the single-particle states and the spatial structure of the
proximity-induced pairing, motivating a systematic study of this correspondence in other quasiperiodic models. The present work is devoted 
to a systematic investigation of the superconducting proximity effect in a 1D hybrid ring, in which the normal region is successively 
replaced by two qualitatively different models of quasiperiodic and topological order. We self-consistently compute the spatial distribution 
of the order parameter using the Bogoliubov-de Gennes (BdG)~\cite{sc50,sc51} framework at zero temperature.

In the first part of this work, the normal region is described by the AAH model. We explore how the proximity-induced pair correlations 
at a given lattice site in the normal region evolve as the quasiperiodic strength is tuned across the localization transition. In the 
extended phase, Cooper pair correlations penetrate deeply into the normal region. As localization sets in, the penetration is progressively
suppressed. This provides a direct, real-space signature of the AAH localization transition in a proximity-effect measurement. 
To characterize the sensitivity of the proximity effect to the specific configuration of the onsite AAH modulation, we analyze the variation 
of the self-consistently determined order parameter $\Delta_{i}$ at a fixed site $i$ in the interior of the normal region for different 
values of the AAH phase $\phi$. Keeping the chain length $L_N$ of the normal region fixed, varying $\phi \in [0, 2\pi]$ generates 
a continuous family of AAH configurations, each characterized by a distinct on-site quasiperiodic potential landscape with fixed quasiperiodic
modulation strength. Although all such configurations generated by varying the phase factor $\phi$ share the same statistical properties in 
the thermodynamic limit, they differ locally in their on-site energy landscapes for any finite $L_N$, leading to sample-to-sample fluctuations
in the proximity-induced pairing amplitude. For every choice of the phase $\phi$, the BdG equations are iteratively solved until a
self-consistent solution is obtained. We then extract the superconducting order parameter at representative sites chosen within the normal
segment. By repeating this procedure for different values of $\phi$, we obtain the phase dependence of $\Delta_{i}$. 
\begin{figure}[ht]
\centering
\includegraphics[width=\columnwidth]{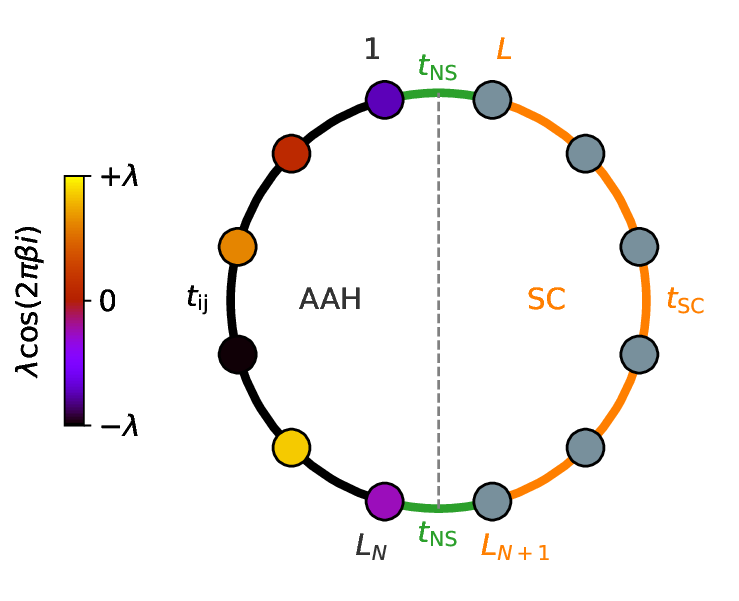}
\caption{(Color online.) Schematic representation of the 1D hybrid ring, consisting of a normal (N) segment of length $L_{N}$ and 
a superconducting (SC) segment of length $L_{SC}=L-L_{N}$. The normal region is characterized by nearest-neighbor hopping amplitude 
$t_{ij}$ together with a quasiperiodic onsite potential of strength $\lambda_{\mathrm{AAH}}$, while the superconducting region is 
described by a uniform hopping amplitude $t_{SC}$. In the normal region, different colored sites correspond to sites with correlated 
potentials, as indicated by the color bar. The N and SC regions are coupled through normal-superconductor interfaces, represented 
by the dashed vertical line, with a hopping amplitude $t_{NS}$.}
\label{fig1}
\end{figure}
This may provide a 
direct measure of how sensitive the proximity-induced superconductivity is to changes in the quasiperiodic potential landscape while keeping 
the quasiperiodic modulation strength $\lambda_{\mathrm{AAH}}$ constant.

We next turn to the normal region described by the SSH-AAH model. In this system, the alternating hopping amplitudes of the SSH chain 
coexist with the quasiperiodic on-site modulation of the AAH model. The SSH-AAH normal region therefore provides a convenient platform 
for studying how the superconducting proximity effect is influenced by both topological band structure and quasiperiodicity. Our analysis 
is carried out in two stages. We begin with $\lambda_{\mathrm{AAH}}=0$, where the normal region is reduced to the SSH chain. In this limit, 
we investigate how the proximity-induced pairing evolves as the system is tuned across the SSH topological phase transition by varying the 
ratio $t_A/t_B$. We then introduce a finite quasiperiodic potential and inspect how it modifies the superconducting correlations established 
in the topological phase. Particular attention is paid to the competition between the robustness associated with hopping dimerization and
the localization tendencies induced by quasiperiodicity. 

The rest part of the work is follows. In Sec.~\ref{sec:model} we describe the Hamiltonian of the hybrid ring, the BdG mean-field 
decoupling method, and the self-consistent numerical procedure for calculating the results. Section~\ref{sec:results} presents all the 
numerical results, organized into three sub-sections, covering the AAH region (\ref{subsec:aah}), the SSH chain (\ref{subsec:ssh}), and 
the combined SSH-AAH model (\ref{subsec:sshaah}). Section~\ref{sec:conclusion} includes the essential findings of our work, an outlook 
for future work, and possible experimental realizations of our proposed model hybrid quantum ring.

\section{Model Hybrid Quantum Ring and the Method}
\label{sec:model}

\subsection{Hamiltonian of the Hybrid Ring}

We consider a hybrid ring of $L$ total sites (see Fig.~\ref{fig1}), consisting of $L_{N}=L-L_{SC}$ normal sites hosting AAH type 
quasiperiodic potential 
and $L_{SC}$ superconducting sites. Throughout this work, the lattice spacing is set to unity, so that the lengths are measured in 
lattice units. The full Hamiltonian of the hybrid ring is written as 
\begin{align}
H = H_{N} + H_{SC} + H_{NS},
\label{eq:full_H}
\end{align}
where  $H_{N}$, $H_{SC}$, and $H_{NS}$ describe the Hamiltonians of the normal, superconducting, and the normal-superconductor (NS) 
interface region, respectively.

\vskip 0.1cm
\noindent 
\textit{Normal region}: The normal region is described by the SSH-AAH tight-binding (TB) Hamiltonian
\begin{align}
H_{N} =
-\sum_{\langle i,j\rangle \in N}
t_{ij}\, c^{\dagger}_{i} c_{j}^{\phantom{\dagger}}
+ \lambda_{\mathrm{AAH}}\sum_{i \in N}
\cos(2\pi\beta i + \phi)\,
c^{\dagger}_{i} c_{i}^{\phantom{\dagger}},
\label{eq:H_N}
\end{align}
where $c^{\dagger}_{i}$ ($c_{i}^{\phantom{\dagger}}$) creates (annihilates) an electron at site $i$, $\lambda_{\mathrm{AAH}}$ is the
quasiperiodic modulation strength, $\beta = (\sqrt{5}-1)/2$ is the inverse golden ratio, and $\phi$ is the AAH phase. The hopping 
amplitude $t_{ij}$ takes the form
\begin{align}
t_{ij} =
\begin{cases}
t_A & \text{if } (i,j) \text{ is an intra-cell bond},\\
t_B & \text{if } (i,j) \text{ is an inter-cell bond}.
\end{cases}
\label{eq:tij}
\end{align}
The Hamiltonian given in Eq.~(\ref{eq:H_N}) represents three different models depending on the choice of the TB parameters. Here,
we represent them one by one as follows.
\begin{itemize}[leftmargin=*, label=\textbullet]
\item \textit{AAH model}: Setting $t_A = t_B \equiv t_N$ makes the hopping uniform throughout
the normal region, and Eq.~(\ref{eq:H_N}) corresponds to the well known AAH model with a
quasiperiodic on-site potential of strength $\lambda_{\mathrm{AAH}}$. 

\item \textit{SSH model}: Setting $\lambda_{\mathrm{AAH}} = 0$ removes the on-site potential entirely, and Eq.~(\ref{eq:H_N}) reduces to 
the SSH chain with alternating bonds $t_A$ and $t_B$. The SSH chain is topologically non-trivial for $t_A < t_B$, whereas it is 
topologically trivial for $t_A > t_B$ with the topological phase transition occurring at $t_A = t_B$ where the bulk gap 
$E_{\mathrm{gap}} = 2|t_B - t_A|$ closes.

\item \textit{SSH-AAH model}: For the general case where $t_A \neq t_B$ and $\lambda_{\mathrm{AAH}} \neq 0$, Eq.~(\ref{eq:H_N}) corresponds
to the SSH-AAH chain, in which the dimerized hopping structure of the SSH chain coexists with quasiperiodic on-site disorder. This model 
allows us to study the competition between hopping dimerization and quasiperiodic localization and their combined effect on the
superconducting proximity effect.
\end{itemize}

The above-mentioned three limiting cases are summarized in Table~\ref{tab:models}.
\begin{table}[ht]
\vskip -0.2cm
\centering
\caption{Three models described by the generalized normal Hamiltonian Eq.~(\ref{eq:H_N}).}
\label{tab:models}
\vskip 0.2cm
\centering
\begin{tabular}{lcc}
\hline\hline
Model & $t_A/t_B$ & $\lambda_{\mathrm{AAH}}$ \\
\hline
AAH       & $t_A = t_B \equiv t_N$ & $\neq 0$ \\
SSH       & $\neq 1$               & $= 0$    \\
SSH--AAH  & $\neq 1$               & $\neq 0$ \\
\hline\hline
\end{tabular}
\end{table}

\vskip 0.2cm
\noindent 
\textit{Superconducting region}: The superconducting region is described by the following Hamiltonian~\cite{sc1}
\begin{align}
H_{SC} = -t_{SC}\!\!\sum_{\substack{i,j \in SC \\
          \langle i,j\rangle,\sigma}}\!\!
          c^{\dagger}_{i\sigma} c_{j\sigma}
        +V_{SC}\!\sum_{i \in SC}
          c^{\dagger}_{i\uparrow} c^{\dagger}_{i\downarrow}
          c_{i\downarrow} c_{i\uparrow},
\label{eq:H_SC}
\end{align}
where $t_{SC}$ is the nearest-neighbor (NN) hopping within the superconductor and $V_{SC} $ is the strength of the local on-site 
attractive interaction between opposite-spin electrons that drives Cooper pairing.

\vskip 0.2cm
\noindent 
\textit{Normal-superconductor interface}: The ring contains two NS interfaces connecting the last site of the normal region to the first 
site of the SC region and the last SC site back to the first normal site. The coupling between the normal and the superconducting regions 
is described by
\begin{align}
H_{NS} = -t_{NS}\!\!\sum_{\langle i,j\rangle_{\mathrm{int}}}\!\!(c^{\dagger}_{i} c_{j}+ h.c.),
\label{eq:H_NS}
\end{align}
where $t_{NS}$ is the hopping amplitude at the NS interfaces. For perfectly transparent interfaces, $t_{NS} = t_{N} = t_{SC}$.

\subsection{The Method}

\subsubsection{Mean-Field BdG Hamiltonian}

The quartic interaction term in $H_{SC}$ is decoupled using the Bogoliubov mean-field approximation~\cite{sc50,sc51,sc52,sc53}. 
We introduce the spatially resolved superconducting order parameter (OP)
\begin{align}
\Delta_{i} \equiv
\langle
c_{i\downarrow}
c_{i\uparrow}
\rangle,
\label{eq:OP_def}
\end{align}
which represents the local pair amplitude at site $i$. Since the pairing interaction is absent in the normal region 
($V_{i} = 0$ for $i \in N$), $V_{i}\Delta_{i}$ is strictly zero there, even though $\Delta_{i}$ itself can be nonzero due to the 
superconducting proximity effect. The interaction also generates a Hartree shift in the on-site energy of the superconducting region which is
\begin{align}
U^{H}_{i} = -V_{i}\sum_{E_{n}<0}|u_{i,n}|^{2},
\label{eq:Hartree}
\end{align}
where $u_{i,n}$ is the particle (electron-like) amplitude at site $i$ of quasiparticle eigenstate $n$ with energy $E_n$, and the sum runs 
over all negative-energy states. Here $V_{i} = V_{SC}$ for $i \in SC$ and $V_{i} = 0$ for $i \in N$, so the Hartree shift is nonzero only 
in the SC region. Combining the AAH potential and the Hartree shift into the total effective onsite energy, we get
\begin{align}
U^{\mathrm{tot}}_{i} =
U^{\mathrm{AAH}}_{i} + U^{H}_{i},
\label{eq:Utot}
\end{align}
where $U^{\mathrm{AAH}}_{i}$ is the on-site AAH potential. The mean-field Hamiltonian reads as
\begin{align}
H_{\mathrm{MF}} &=
\sum_{i,\sigma} U^{\mathrm{tot}}_{i}\,
  c^{\dagger}_{i\sigma} c_{i\sigma}
- \sum_{\langle i,j\rangle,\sigma} t_{ij}\,
  c^{\dagger}_{i\sigma} c_{j\sigma}
\nonumber\\
&\quad
+\sum_{i \in SC}
  \bigl(
  V_{i}\Delta_{i}\,
  c^{\dagger}_{i\uparrow} c^{\dagger}_{i\downarrow}
  + h.c.
  \bigr),
\label{eq:H_MF}
\end{align}
where $t_{ij}$ takes the value $t_{A}$, $t_{B}$, $t_{SC}$, or $t_{NS}$ according to whether the bond $(i,j)$ lies entirely within the 
normal region, entirely within the SC region, or crosses an NS interface, respectively.

\subsubsection{Bogoliubov-de Gennes method}

The Hamiltonian $H_{\mathrm{MF}}$ is diagonalized via the Bogoliubov transformation
\begin{align}
c_{i\uparrow} &= \sum_{n}\bigl(
  u_{i,n}\,\gamma_{n\uparrow}
- v^{*}_{i,n}\,\gamma^{\dagger}_{n\downarrow}\bigr),
\nonumber\\
c_{i\downarrow} &= \sum_{n}\bigl(
  u_{i,n}\,\gamma_{n\downarrow}
+ v^{*}_{i,n}\,\gamma^{\dagger}_{n\uparrow}\bigr),
\label{eq:Bog_transform}
\end{align}
where $\gamma^{\dagger}_{n\sigma}$ ($\gamma^{\phantom{\dagger}}_{n\sigma}$) creates (annihilates) a Bogoliubov quasiparticle in state 
$n$ with spin $\sigma$ ($\uparrow,\downarrow$), and $u_{i,n}$, $v_{i,n}$ are the particle and hole amplitudes at site $i$ in eigenstate 
$n$. Expressing $H_{\mathrm{MF}}$ in the Nambu spinor basis
$\Psi = (c_{1\uparrow},\ldots,c_{L\uparrow},
         c^{\dagger}_{1\downarrow},\ldots,
         c^{\dagger}_{L\downarrow})^{T}$
yields the $(2L\times 2L)$ BdG matrix
\begin{align}
\mathcal{H}_{\mathrm{BdG}} =
\begin{pmatrix}
  \hat{K} & \hat{\Delta} \\[4pt]
  \hat{\Delta}^{\dagger} & -\hat{K}
\end{pmatrix},
\label{eq:BdG_matrix}
\end{align}
where the ($L\times L$) blocks are defined
element-wise as
\begin{align}
K_{ij} &= U^{\mathrm{tot}}_{i}\,\delta_{ij}
          - t_{ij}(1-\delta_{ij}),
\label{eq:K_block}
\\
\Delta_{ij} &= V_{i}\,\Delta_{i}\,\delta_{ij}.
\label{eq:Delta_block}
\end{align}
Here $\delta_{ij}$ is the Kronecker delta. The diagonal of $\hat{K}$ contains the effective on-site energies $U^{\mathrm{tot}}_{i}$, 
while its off-diagonal elements encode the NN hopping $t_{ij}$ between neighboring sites. The pairing block $\hat{\Delta}$ is diagonal and nonzero only within the SC region, where it equals $V_{SC}\Delta_{i}$, and it is identically zero in the
normal region because $V_{i}=0$ there. The particle-hole symmetry of the BdG formalism is
reflected in the hole block $-\hat{K}$, whose sign is opposite to the particle block. The BdG equation
\begin{align}
\mathcal{H}_{\mathrm{BdG}}
\begin{pmatrix} \bm{u}_{n} \\ \bm{v}_{n} \end{pmatrix}
= E_{n}
\begin{pmatrix} \bm{u}_{n} \\ \bm{v}_{n} \end{pmatrix}
\label{eq:BdG_eig}
\end{align}
yields $2L$ eigenpairs $\{E_{n},(\bm{u}_{n},\bm{v}_{n})\}$, where $\bm{u}_{n} = (u_{1,n},\ldots,u_{L,n})^{T}$ and $\bm{v}_{n} = (v_{1,n},\ldots,v_{L,n})^{T}$. Particle-hole symmetry guarantees that eigenvalues appear in $\pm E_{n}$ pairs, it suffices to sum
over states with $E_{n} < 0$ in the self-consistency equations below.

\begin{figure*}[t]
\centering
\includegraphics[width=\textwidth]{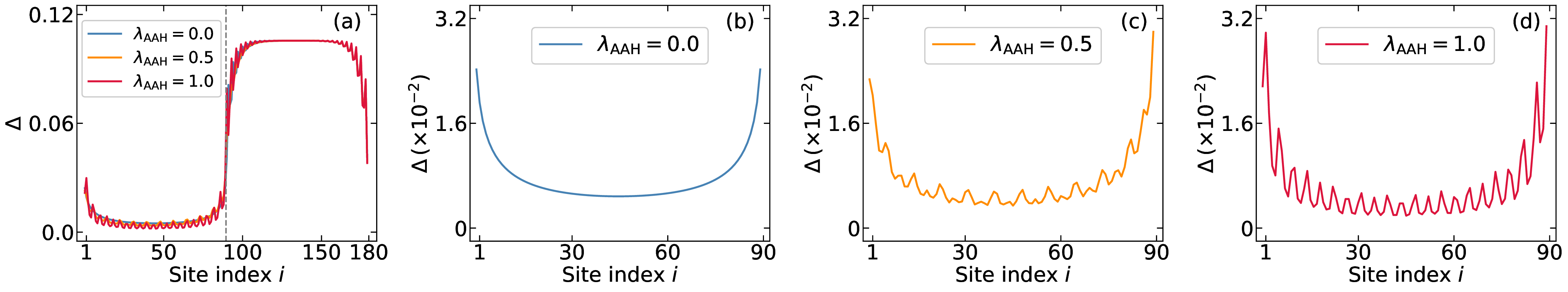}
\caption{(Color online.) Spatial profile of the self-consistently determined pairing amplitude $\Delta_i$ as a function of the site 
index $i$ for three distinct values of the quasiperiodic potential strength $\lambda_{\mathrm{AAH}}$. Panel (a) shows the order parameter 
over the entire hybrid ring for $\lambda_{\mathrm{AAH}}=0$, $0.5$, and $1$. Panels (b), (c), and (d) present enlarged views of the normal 
region for $\lambda_{\mathrm{AAH}}=0$, $0.5$, and $1$, respectively, highlighting the evolution of the spatial fluctuations of the induced
pairing amplitude with increasing quasiperiodic potential strength. The dashed vertical line indicates the normal-superconductor interface. 
The system parameters are $L=180$, $L_N=90$, $L_{SC}=90$, $t_N=t_{SC}=t_{NS}=1$, $V_{SC}=1.5$, and $\phi=0$.}
\label{fig2}
\end{figure*}

\subsubsection{Self-Consistency equations}
\label{sec:selfcon}

The OP $\Delta_{i}$ and the Hartree shift $U^{H}_{i}$ are obtained through an iterative self-consistent procedure~\cite{sc8,sc9}. Starting from an initial guess $\Delta^{(0)}_{i} = 1$ for $i \in SC$, $\Delta^{(0)}_{i} = 0$ for $i \in N$, and $U^{H,(0)}_{i} = 0$ everywhere, at each iteration $k$ we diagonalize $\mathcal{H}_{\mathrm{BdG}}$ and update the OP and Hartree shift as
\begin{align}
\Delta^{(k+1)}_{i} &=
  -\sum_{E_{n}<0} u_{i,n}\,v^{*}_{i,n},
\label{eq:Delta_selfcon}
\\
U^{H,(k+1)}_{i} &=
  -V_{i}\sum_{E_{n}<0}|u_{i,n}|^{2},
\label{eq:UH_selfcon}
\end{align}
where the sums are taken over all quasiparticle eigenstates with $E_{n} < 0$, computed using the values $\Delta^{(k)}_{i}$ and 
$U^{H,(k)}_{i}$ from the previous step. This procedure is repeated until the solution converges, as described below.

\subsubsection{Convergence}

We monitor convergence by checking the relative change in $\Delta_{i}$ between successive
iterations~\cite{sc8,sc9}. At each site $i$, we set
\begin{align}
\frac{|\Delta^{(k+1)}_{i} - \Delta^{(k)}_{i}|}
     {|\Delta^{(k+1)}_{i}|}
< 10^{-3}.
\label{eq:convergence}
\end{align}
Once Eq.~(\ref{eq:convergence})
holds at every relevant site, the loop is stopped and the converged $|\Delta_{i}|$ is taken as the physical result.

\section{Numerical Results and Discussion}
\label{sec:results}

\subsection{AAH Normal Region}
\label{subsec:aah}
We express all energies in units of electron-volt (eV) throughput the work. The hopping amplitudes are set to $t_N=t_B=t_{SC}=t_{NS}=1$ 
and the attractive Hubbard interaction strength is taken as $V_{SC}=1.5$.

To gain insight into the proximity effect within the AAH region, we first explore the spatial distribution of the self-consistently 
determined pairing amplitude. Figure~\ref{fig2} shows the real-space profile of OP $\Delta_i$ as a function of site index $i$ for 
three different values of the AAH potential strength $\lambda_{\mathrm{AAH}} = 0, 0.5, 1$, all well within the extended phase
($\lambda_{\mathrm{AAH}} < 2t_N = 2$). The spatial variation of the order parameter in the normal region is shown separately in
Figs.~\ref{fig2}(b)-(d) for the mentioned three values of the quasiperiodic modulation strength, so that the spatial fluctuations 
within each region can be inspected simultaneously without the curves overlapping. We discuss the behavior in each region separately.

In the SC region, the order parameter attains a large, approximately uniform bulk value in the interior of the SC chain, reflecting 
the self-consistent pairing driven by the attractive on-site interaction $V_{SC}$. The bulk value is determined primarily by the 
pairing interaction $V_{SC}$ and the density of states of the SC region, and remains essentially unchanged across the three values 
of $\lambda_{\mathrm{AAH}}$ studied here, since the AAH potential acts only in the normal region and does not directly affect the 
SC pairing in the bulk. Close to the NS interfaces, however, the induced pairing amplitude is visibly reduced from its bulk value. 
This suppression originates from the inverse proximity effect, where electrons from the normal region penetrate into the superconductor 
and locally weaken the pairing correlations. Superimposed on the suppression near the interfaces, small oscillations in $\Delta_i$ are 
visible, reminiscent of Friedel oscillations arising from the interference of quasiparticle states reflected at the NS interface. 
These features are present for all three values of $\lambda_{\mathrm{AAH}}$ and become slightly more pronounced as 
$\lambda_{\mathrm{AAH}}$ increases, consistent with a stronger electronic mismatch between the normal and SC regions at larger 
quasiperiodic potential strengths.

In the normal region, the order parameter $\Delta_i$ is proximity-induced. Since the pairing interaction is absent ($V_i = 0$) 
in this region, $V_i\Delta_i$ becomes zero, and the nonzero values of $\Delta_i$ reflect the pairing amplitude 
$\langle c_{i\downarrow} c_{i\uparrow}\rangle$ leaking from the SC through the NS interfaces via the Andreev processes. The 
proximity-induced pairing amplitude within the AAH region is significantly smaller in magnitude than the bulk value in SC region, 
as expected. The spatial profile of $\Delta_i$ in the normal region is not smooth but exhibits site dependent fluctuations, whose 
amplitude increases with $\lambda_{\mathrm{AAH}}$. These fluctuations arise from the quasiperiodic modulation of the onsite potential, 
which modifies the spatial structure of the single-particle wave functions and, consequently the local pairing amplitude. For $\lambda_{\mathrm{AAH}} = 0$ (blue curve) the profile is nearly smooth, consistent with extended plane-wave-like eigenstates. 
As $\lambda_{\mathrm{AAH}}$ increases (orange and red curves), the stronger quasiperiodic modulation of the wave functions leads 
to more pronounced spatial variations in pairing amplitude.
\begin{figure*}[t]
\centering
\includegraphics[width=\textwidth]{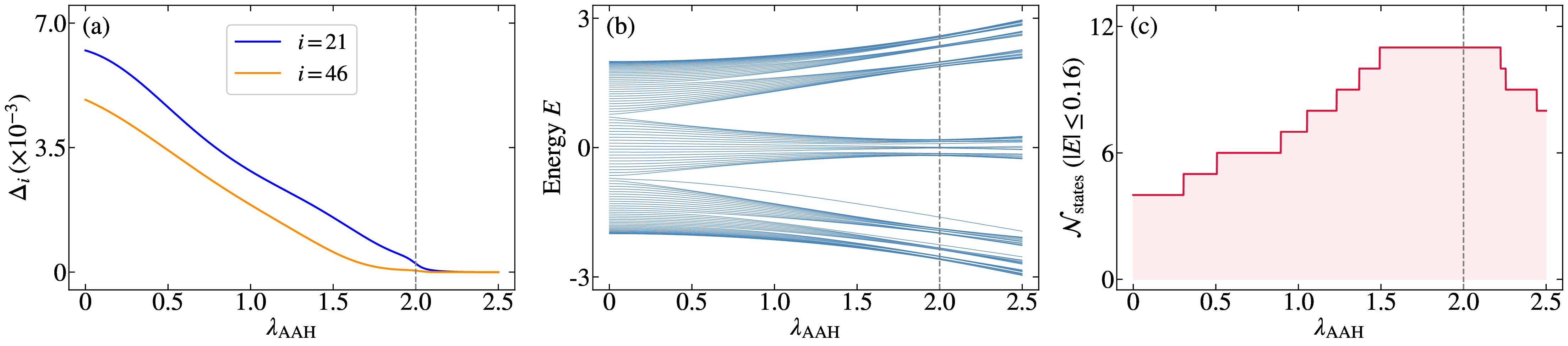}
\caption{(Color online.) (a) Self-consistently determined superconducting order parameter $\Delta_i$ at two representative sites 
($i=21$ and $i=46$) in the normal region as a function of the quasiperiodic potential strength $\lambda_{\mathrm{AAH}}$. The vertical 
dashed line at $\lambda_{\mathrm{AAH}}=2t_N$ marks the localization transition of the AAH chain. (b) Energy spectrum of the finite AAH 
chain with $L_N=90$ sites as a function of $\lambda_{\mathrm{AAH}}$, illustrating the evolution of the eigenstates across the localization
transition. (c) Number of quasiparticle states, $\mathcal{N}_{\mathrm{states}}$, with energies lying within the window $[-0.16,\,0.16]$ 
as a function of $\lambda_{\mathrm{AAH}}$. The remaining system parameters are the same as in Fig.~\ref{fig2}.}
\label{fig3}
\end{figure*}
At the same time, the overall magnitude of the induced order parameter gradually decreases as $\lambda_{\rm AAH}$ is increased. The
quasiperiodic potential progressively reduces the ability of Cooper pairs to propagate through the normal region by suppressing the 
spatial extent of the single particle states  participating in the proximity effect. Consequently, superconducting correlations penetrate 
less efficiently into the AAH segment, leading to a reduction of the induced pair amplitude throughout the normal region. Therefore, 
increasing quasiperiodicity has a dual effect. It enhances the spatial inhomogeneity of the induced superconducting correlations 
while simultaneously reducing their strength. These real-space characteristics provide the microscopic origin of the behavior observed 
in the subsequent analysis of the induced order parameter as a function of the quasiperiodic modulation strength.

To quantify this suppression and relate it to the localization transition of the AAH chain we analyze the order parameter $\Delta_{i}$ 
at two different lattice sites ($i = 21, 46$) in the normal region as a function of $\lambda_{\mathrm{AAH}}$. One can see from
Fig.~\ref{fig3}(a) that $\Delta_{i}$ decreases monotonically as the quasiperiodic modulation strength is increased, and drops to essentially
to zero once $\lambda_{\mathrm{AAH}}$ crosses the critical value $2t_N$. This is a direct indication that the proximity effect is completely
diminished once the normal-region states become localized since the leakage of the Cooper-pair into the normal region reduces as
$\lambda_{\mathrm{AAH}}$ drives the system from extended to the localized region. To understand the mechanism behind this suppression, 
we plot the energy eigenvalues as a function of $\lambda_{\mathrm{AAH}}$ in Fig.~\ref{fig3}(b), which reflects the progressive fragmentation 
of the AAH spectrum from a continuous band in the extended phase to a fractal-like structure at the critical point, as 
$\lambda_{\mathrm{AAH}} \to 2t_N$. We further investigate the consequences of the spectral restructuring on the induced order parameter, 
and for that we compute the number of states $\mathcal{N}_{\mathrm{states}}$ with energies in the window $[-0.16, 0.16]$ as a function 
of $\lambda_{\mathrm{AAH}}$. The results is shown in Fig.~\ref{fig3}(c). This energy window corresponds directly to the energy scale
$V_{SC}\Delta_{mid}$, where $\Delta_{mid}$ is the self-consistently determined pairing amplitude evaluated at the center of the 
SC region. This allows us to evaluate how many states are available to participate in the pairing. Remarkably, 
$\mathcal{N}_{\mathrm{states}}$ increases steadily with $\lambda_{\mathrm{AAH}}$ and saturates in the vicinity of the critical point 
capturing dense accumulation of the multifractal states. For $\lambda_{\mathrm{AAH}} > 2t_N$, however, $\mathcal{N}_{\mathrm{states}}$ 
decreases again as the system enters the localized phase. In this regime, the eigenstates become exponentially localized, reducing 
their spatial overlap and thereby limiting their ability to support proximity-induced pairing correlations.

The simultaneous vanishing of $\Delta_{i}$ and growth of $\mathcal{N}_{\mathrm{states}}$ establishes that the suppression of the 
proximity effect is {\em not} due to a depletion of states around the Fermi energy. Rather, it is a consequence of the \textit{spatial}
localization of these states. If the density of states were the controlling factor, one would expect a larger number of states to enhance 
the induced pairing correlations. The opposite trend is observed here. Although more states accumulate within the superconducting energy 
window as $\lambda_{\rm AAH}$ increases, their ability to participate in the proximity effect is progressively reduced.
The reason lies in the localization properties of the eigenstates. The proximity-induced pairing amplitude is governed by the 
self-consistency equation described in Eq.~(\ref{eq:Delta_selfcon}) in which the contribution of each quasiparticle state $n$ is weighted 
by the product of its particle and hole amplitudes $u_{i,n}$ and $v^*_{i,n}$ at site $i$. In the localized phase, eigenstates are 
exponentially confined with localization length $\xi \sim 1/\ln(\lambda_{\mathrm{AAH}}/2t_N)$~\cite{sc18}, so that states localized away 
from site $i$ contribute negligibly to $\Delta_{i}$ regardless of their energy. The growing number of sub-gap states therefore fails to 
enhance the proximity effect because these states have exponentially small amplitude at $i$. The suppression of $\Delta_{i}$ thus reflects 
the dominance of spatial localization over the spectral accumulation of low-energy states, and constitutes a clear real-space, 
self-consistent BdG signature of the AAH localization-delocalization transition. 
 
To characterize the sensitivity of the proximity effect to the specific configurations of the quasiperiodic potential, we track the
self-consistently 
\begin{figure}[ht]
\centering 
\includegraphics[width=0.475\textwidth]{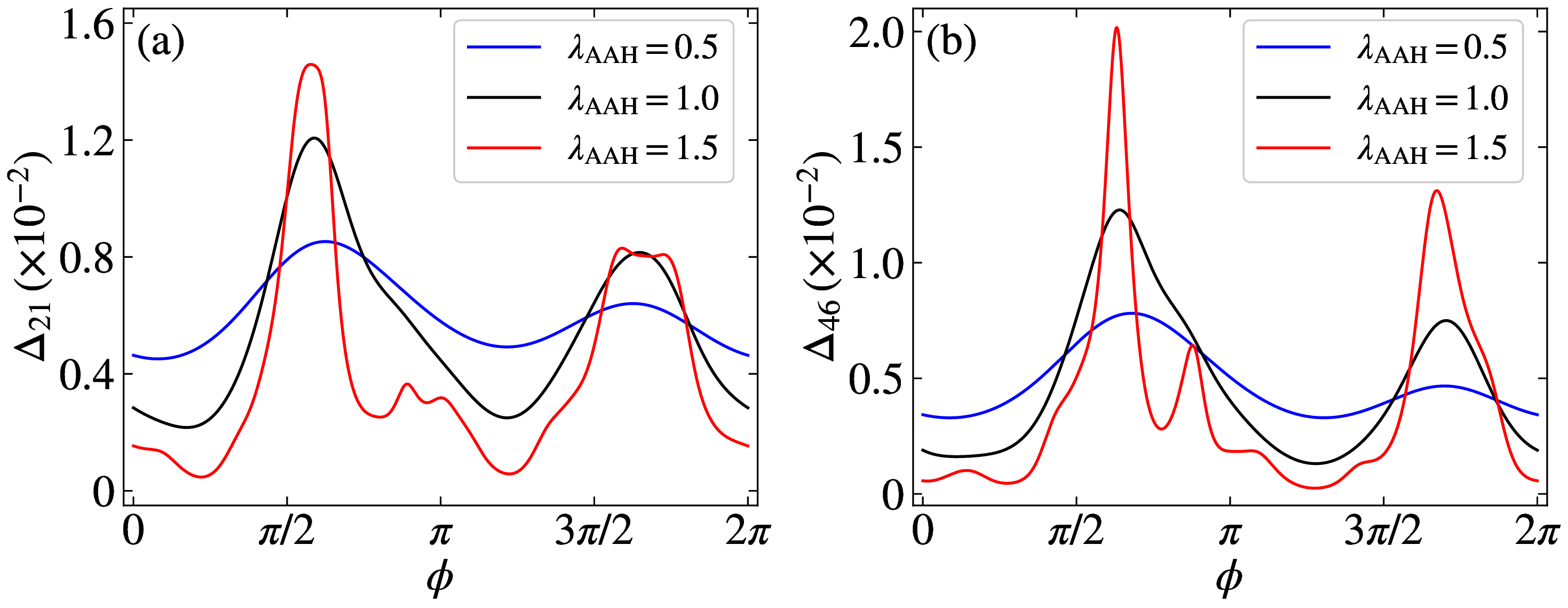}
\caption{(Color online.) Variation of the order parameter $\Delta_i$ at sites (a) $i = 21$  and (b) $i = 46$ as a function of the AAH 
phase $\phi \in [0, 2\pi]$ for three distinct values of the quasiperiodic potential strength $\lambda_{\mathrm{AAH}} = 0.5$, $1.0$, and 
$1.5$. The remaining system parameters are the same as in Fig.~\ref{fig2}.}
\label{fig4}
\end{figure}
\begin{figure*}[t]
\centering
\includegraphics[width=\textwidth]{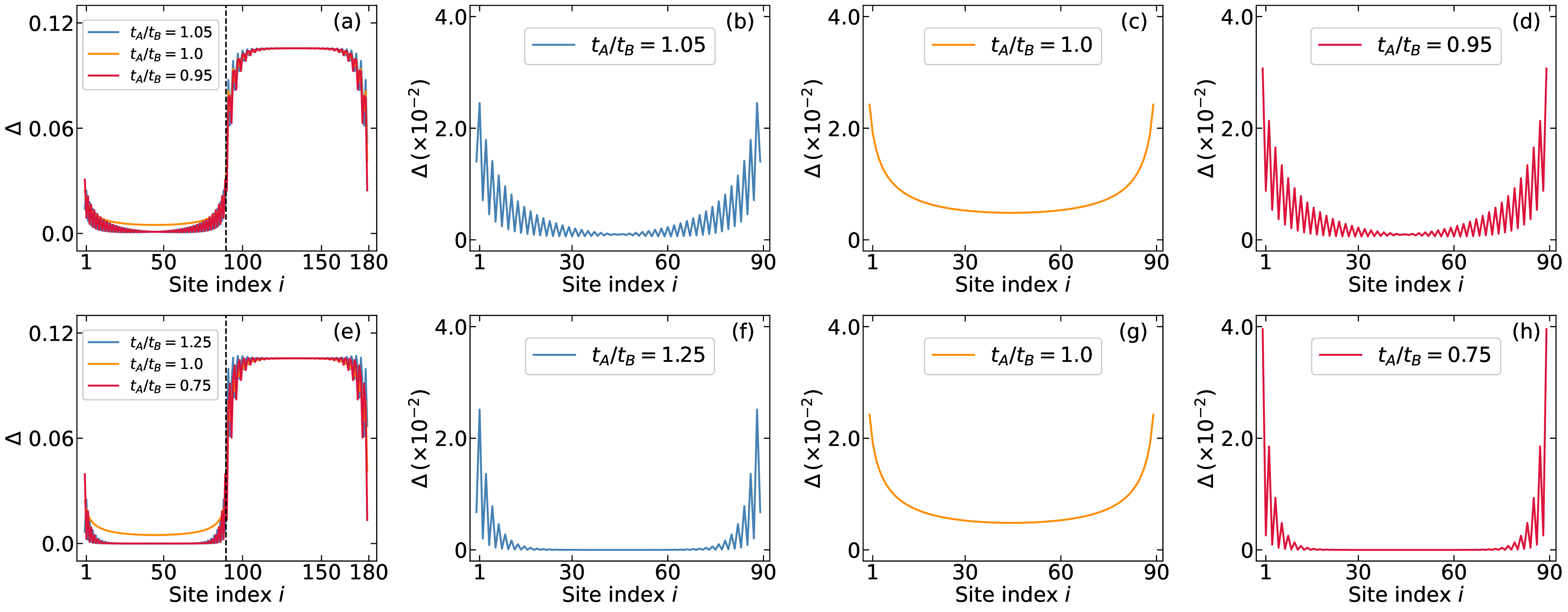}
\caption{(Color online.) Real-space profile of the self-consistently determined superconducting pairing amplitude $\Delta_i$ as a function 
of the site index $i$ for different values of the SSH dimerization ratio $t_A/t_B$. Panels (a) and (e) show the variation of the order 
parameter over the entire hybrid ring for the weakly and strongly dimerized cases, respectively. Panels (b) and (d) display enlarged views 
of the normal region for the weakly dimerized cases with $t_A/t_B=1.05$ and $0.95$. Similarly, panels (f) and (h) show the strongly dimerized
cases with $t_A/t_B=1.25$ and $0.75$. Whereas panel (c) and (g) represent the uniform chain ($t_A/t_B=1$) for comparison. The dashed vertical
line indicates the normal-superconductor interface. The system parameters are $L=180$, $L_N=90$, $L_{SC}=90$, $t_{SC}=t_{NS}=1.0$, and
$V_{SC}=1.5$.}
\label{fig5}
\end{figure*}
determined order parameter at two sites $i=21$ and $46$ as the AAH phase $\phi$ is varied which is shown in Figs.~\ref{fig4}(a) and (b)
respectively. For a chain of fixed length $L_N$, sweeping $\phi \in [0, 2\pi]$ generates a continuous family of quasiperiodic potentials 
that are statistically equivalent in the thermodynamic limit but differ locally in their on-site energy landscapes for any finite $L_N$. 
Thus, varying $\phi$ continuously samples different configurations of the quasiperiodic potential, and the resulting curve $\Delta_{i}(\phi)$
directly captures the sample-to-sample fluctuations of the proximity-induced pairing amplitude at fixed disorder strength $\lambda_{\mathrm{AAH}}$.
For all the three cases, $\Delta_{i}(\phi)$ fluctuates smoothly with $\phi$, which is a direct signature of the finite-size effect. 
Notably, the amplitude of these fluctuations grows with increasing $\lambda_{\mathrm{AAH}}$, indicating that the proximity-induced pairing
amplitude becomes increasingly sensitive to the specific configuration of the quasiperiodic potential as the quasiperiodic modulation 
strength increased. While the fluctuations across different values of $\phi$ remain relatively modest in the weak to moderate quasiperiodic
regime, they become substantially more pronounced at stronger disorder, reflecting a growing dependence of the local pairing amplitude on 
the fine details of the on-site energy landscape.

\subsection{SSH Normal Region}
\label{subsec:ssh}

Having studied the AAH case, we now turn to the SSH model and investigate its impact on the proximity-induced superconducting correlations. 
The spatial distribution of the self-consistent order parameter $\Delta_{i}$ for different hopping ratios $t_A/t_B$ is shown in 
Fig.~\ref{fig5}. Panels (a) and (e) of Fig.~\ref{fig5} display the spatial variation of the order parameter over the entire hybrid ring 
for the weakly and strongly dimerized cases, respectively. The corresponding enlarged views shown in Figs.~\ref{fig5} (b)-(d) and (f)-(h) 
focus on the normal segment, making the evolution of the proximity-induced pairing correlations easier to visualize.

For the uniform chain, $t_A/t_B=1$, the order parameter decreases smoothly from the NS interfaces towards the middle of the normal region.
Although the induced pairing is weakest near the center, it remains finite throughout the chain, indicating that superconducting correlations
can penetrate relatively far into the normal segment.
\begin{figure*}[t]
\centering
\includegraphics[width=\textwidth]{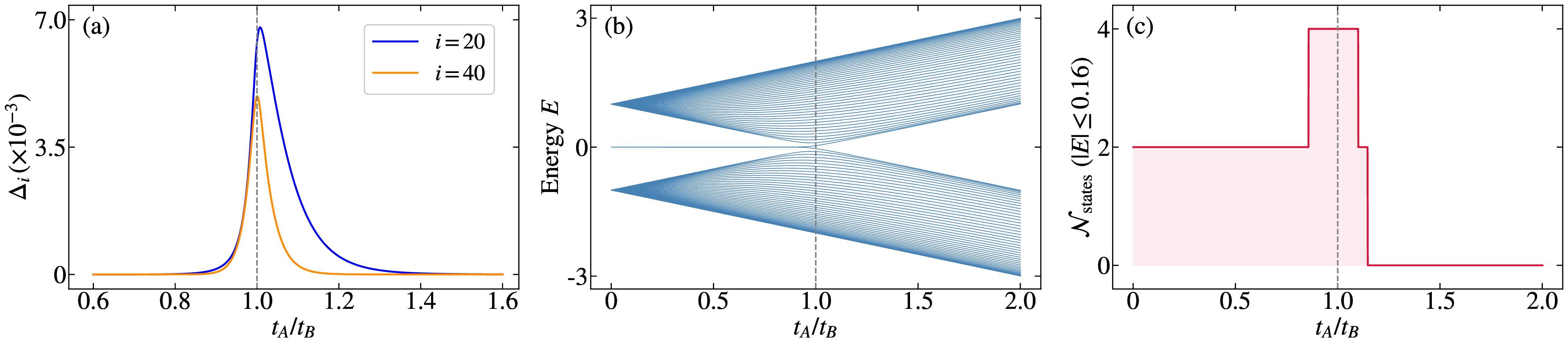}
\caption{(Color online.) (a) Self-consistently determined superconducting order parameter $\Delta_i$ at two representative sites 
($i=20$ and $i=40$) in the normal region as a function of the SSH dimerization ratio $t_A/t_B$. The induced pairing reaches its maximum 
near $t_A/t_B=1.0$, where the SSH chain undergoes a topological phase transition. (b) Energy spectrum of the finite SSH chain with $L_N=90$
sites as a function of the dimerization ratio $t_A/t_B$, showing the evolution of the bulk bands and the gap closing at $t_A/t_B=1$. 
(c) Number of states, $\mathcal{N}_{\mathrm{states}}$, with energies lying within the window $[-0.16,\,0.16]$ as a function of $t_A/t_B$. 
The remaining system parameters are the same as in Fig.~\ref{fig5}.}
\label{fig6}
\end{figure*}

As soon as a small dimerization is introduced, the smooth profile gives way to an oscillatory one. From Figs.~\ref{fig5}(b)-(d) we can 
see that in the weak dimerization case for both $t_A/t_B=1.05$ and $0.95$, where clear oscillations appear near the interfaces and continue 
well into the interior of the chain. The overall envelope of the induced pairing is not very different from the uniform case, but its spatial
modulation is noticeably modified. The oscillations survive over a large part of the normal region before gradually diminishing towards the
center, suggesting that weak dimerization alters the distribution of the superconducting correlations without strongly suppressing them.
A stronger effect is observed for $t_A/t_B=1.25$ and $0.75$ which is shown in Figs.~\ref{fig5}(f)-(h). In these cases, the oscillations are
concentrated near the interfaces and decay much more rapidly with distance. As a result, the induced pairing becomes extremely small in the
middle of the chain. Compared with the weakly dimerized cases, the range over which superconducting correlations extend into the normal 
region is significantly reduced. The proximity effect therefore becomes increasingly confined to the vicinity of the NS boundaries as the
hopping imbalance is increased. In this regime, the SSH chain behaves as a poor medium for Cooper pair leakage from the SC region. The 
strong alternation between $t_A$ and $t_B$ effectively suppresses transport through the bulk, leading to a reduced penetration depth of
superconducting correlations. 

To get a more quantitative handle on how dimerization influences the proximity effect, we fix two sites in the normal region, $i = 20$
and $i = 40$, and track the induced order parameter $\Delta_{i}$ at those sites as we sweep the dimerization ratio $t_A/t_B$. The
result is shown in Fig.~\ref{fig6}(a). What stands out immediately is how sharply $\Delta_{i}$ peaks at $t_A/t_B = 1$ and how quickly 
it falls away on either side. The site closer to the NS interface ($i = 20$) carries a larger induced pairing than the deeper site 
($i = 40$), as one would expect, but both sites tell the same story. This behavior is a direct consequence of the bulk energy gap of the 
SSH chain defined as
\begin{align}
E_{\mathrm{gap}} = 2t_B\left|1 - \frac{t_A}{t_B}\right|,
\label{eq:SSH_gap}
\end{align}
which closes at $t_A/t_B = 1$ and grows linearly on either side which can be seen from Fig.~\ref{fig6}(b). The proximity-induced order 
parameter is dominated by quasiparticle states near the Fermi level, which carry the largest mixed particle-hole character as measured 
by the product $u_{i,n}v^*_{i,n}$ in the self-consistency equation. When $t_A/t_B = 1$, the SSH energy spectrum is gapless which means 
there are large number of energy labels available near the energy which can also be seen from Fig.~\ref{fig6}(c) that within the low-energy
window $|E| \leq 0.16$ the count hits its maximum. These are the states which actually do the work for carrying the Cooper pairs from SC 
region to the SSH region. So having more of them naturally makes the proximity effect stronger. The moment $t_A/t_B$ moves away from unity, 
the spectrum becomes gapped and the number of states within that energy windows drops off with the gap opening shown in the Fig.~\ref{fig6}(b).
It does not open gradually, it grows quite quickly as the dimerization strengthens, and it leaves fewer and fewer states available to mediate
the pairing which can also be seen from Fig.~\ref{fig6}(c). So, when the SSH gap is large, no such states exist within the relevant energy
window and the induced superconducting correlations is suppressed. The observed variation of $\Delta_{i}$ is therefore serves as a real-space,
self-consistent BdG signature of the SSH topological phase transition.

\begin{figure*}[t]
\centering
\includegraphics[width=\textwidth]{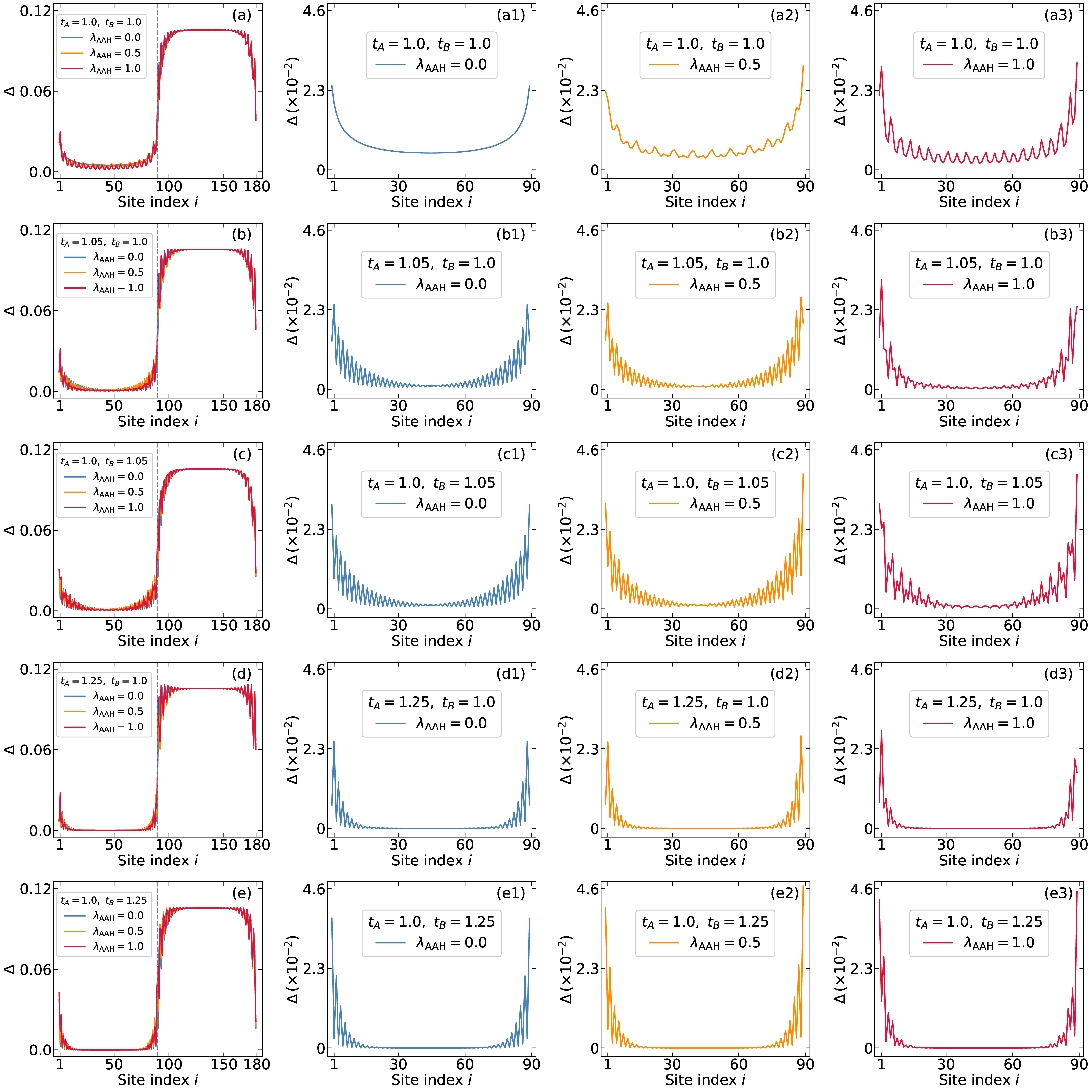}
\caption{(Color online.) Self-consistently determined real-space profile of the pairing amplitude $\Delta_i$ as a function of the site 
index $i$ for different combinations of the SSH dimerization ratio $t_A/t_B$ and the AAH quasiperiodic potential strength
$\lambda_{\mathrm{AAH}}$. In all cases, results are shown for $\lambda_{\mathrm{AAH}}=0$, $0.5$, and $1$. Panels (a)-(e) display the 
spatial variation of $\Delta_i$ over the full hybrid ring for the uniform chain ($t_A/t_B=1$), the weakly dimerized cases ($t_A/t_B=1.05$ 
and $0.95$), and the strongly dimerized cases ($t_A/t_B=1.25$ and $0.75$), respectively. Panels (a1)-(a3) show enlarged views of the 
normal region for the uniform chain. Panels (b1)-(b3) and (c1)-(c3) correspond to the weakly dimerized cases with $t_A>t_B$ and $t_A<t_B$,
respectively, while panels (d1)-(d3) and (e1)-(e3) correspond to the strongly dimerized cases with $t_A>t_B$ and $t_A<t_B$, respectively. 
The dashed vertical lines mark the NS interfaces. The system parameters are $L=180$, $L_N=90$, $L_{SC}=90$, $t_{SC}=t_{NS}=1$, $V_{SC}=1.5$, 
and $\phi=0$.}
\label{fig7}
\end{figure*}

\subsection{SSH-AAH Normal Region}
\label{subsec:sshaah}

We now inspect the combined effect of the SSH dimerization and the AAH quasiperiodic potential on the proximity-induced pairing correlations.
The spatial variations of the self-consistent order parameter for several SSH configurations and different values of $\lambda_{\mathrm{AAH}}$ 
is shown in Fig.~\ref{fig7}. Panels (a)-(e) show the order parameter over the entire hybrid ring, while Figs.~\ref{fig7}(a1)-(a3), (b1)-(b3),
(c1)-(c3), (d1)-(d3), and (e1)-(e3) provide enlarged views of the normal segment. Within each set of enlarged panels, the results correspond 
to $\lambda_{\mathrm{AAH}}=0$, $0.5$, and $1$, respectively.

The first row shown in Figs.~\ref{fig7}(a1)-(a3) corresponds to the uniform chain ($t_A=t_B=1$). In the absence of the AAH potential, the
induced order parameter varies smoothly across the normal region and remains finite throughout its length. As $\lambda_{\mathrm{AAH}}$ is
increased, this smooth profile gradually develops pronounced site-to-site variations. Although superconducting correlations still penetrate 
the entire normal segment, their magnitude becomes increasingly nonuniform, and the induced pairing amplitude in the bulk decreases.

The weakly dimerized chains represent an intermediate regime between the uniform and strongly dimerized limits are shown in
Figs.~\ref{fig7}(b1)-(b3) for $t_A/t_B=1.05$ and Figs.~\ref{fig7}(c1)-(c3) for $t_A/t_B=0.95$. In both cases the induced order parameter
exhibits clear oscillations that originate near the normal-superconductor interfaces and persist over a large part of the normal region. 
Unlike the strongly dimerized case, these oscillations are not confined to the interface but extend deep into the bulk, indicating that
superconducting correlations can still propagate over long distances. As the quasiperiodic potential strength $\lambda_{\mathrm{AAH}}$ 
is increased, the oscillatory pattern becomes increasingly irregular and its amplitude is gradually suppressed. Consequently, the combined
effect of weak dimerization and quasiperiodic modulation leads to a reduction in both the coherence and the penetration depth of the
proximity-induced superconducting correlations.

A distinctly different behavior is observed for the strongly dimerized chains which is shown in Figs.~\ref{fig7}(d1)-(d3) and (e1)-(e3). 
For $t_A=1.25,t_B=1$ and $t_A=1,t_B=1.25$, the induced order parameter exhibits pronounced oscillations only in the vicinity of the NS
interfaces. Unlike the weakly dimerized case, these oscillations decay rapidly and do not extend into the bulk of the normal region, 
leaving only a very small pairing amplitude near the center of the chain. Upon introducing the quasiperiodic AAH potential, the oscillatory
pattern close to the interfaces becomes more irregular, while the residual superconducting correlations in the bulk are further suppressed.
Since the penetration of Cooper pairs is already strongly limited by the large hopping imbalance, the overall reduction of the proximity 
effect in this regime is governed primarily by the SSH dimerization, with the quasiperiodic potential providing an additional but 
comparatively weaker suppression.
The degree of dimerization primarily determines how far the proximity-induced pairing correlations can penetrate into the normal region and how they oscillate spatially, while the quasiperiodic potential mainly introduces spatial fluctuations and suppresses the magnitude of the induced pairing. When both are present, the resulting order-parameter profile reflects the competition between these two tendencies, leading to increasingly superconducting correlations as either the dimerization strength or $\lambda_{\mathrm{AAH}}$ is increased.

\section{Closing Remarks and Outlook}
\label{sec:conclusion}

The present work explores the superconducting proximity effect in a 1D hybrid ring by considering different realizations of the normal 
region within a self-consistent Bogoliubov-de Gennes framework. By studying the diagonal AAH model, SSH model, and the combined SSH-AAH 
model, we investigate how quasiperiodicity and hopping dimerization influence the propagation of superconducting correlations into the 
normal segment.

For the diagonal AAH model, increasing the quasiperiodic potential strength gradually suppresses the induced order parameter in the 
normal region while simultaneously enhancing its spatial fluctuations. The AAH phase dependence of the order parameter shows that 
different quasiperiodic realizations can exhibit noticeable variations in the induced pairing for finite systems.

For the SSH model, the dimerization ratio plays a vital role in determining the spatial profile of the induced pairing correlations. 
In the uniform limit, the proximity-induced pairing extends throughout the normal region with a smooth spatial variation. The weak 
dimerization introduces oscillatory modulations in the order parameter, and these oscillations persist over a large portion of the 
normal region, indicating that superconducting correlations can still penetrate deep into the bulk. As the dimerization strength increases, 
the oscillations become confined to regions near the normal-superconductor interfaces and decay rapidly away from the boundaries. 
Consequently, the penetration depth of the proximity effect is significantly reduced, leaving only a negligible induced pairing in the 
central part of the chain. The topological phase transition of the SSH chain also reflected in the behavior of proximity induced pairing.

The combined SSH-AAH model reveals the interplay between hopping dimerization and quasiperiodic modulation. The SSH hopping pattern 
determines the overall oscillatory character and penetration of the induced pairing, while the AAH potential introduces additional 
spatial fluctuations and further suppresses its magnitude. In the weakly dimerized regime, the oscillatory pairing profile survives 
deep inside the normal region but becomes increasingly irregular as the quasiperiodic potential strength is increased. In contrast, 
for strong dimerization, where the induced pairing is already localized near the interfaces, the AAH potential mainly enhances the 
suppression of the residual superconducting correlations.

Overall, the present results demonstrate that hopping dimerization and quasiperiodicity provide two complementary mechanisms for 
controlling proximity-induced superconductivity in one-dimensional hybrid systems. While dimerization governs the penetration depth 
and oscillatory nature of the induced pairing, the quasiperiodic potential controls its spatial inhomogeneity and overall magnitude. 
Their combined effect offers a flexible route for engineering superconducting correlations in low-dimensional quasiperiodic structures 
and may be useful for designing hybrid devices with tunable superconducting properties. Future investigations may explore the effects 
of finite temperature, and longer-range hopping. It would also be interesting to study circular charge and spin currents by applying 
magnetic flux through the hybrid ring.

The experimental realization of quasiperiodic systems has advanced considerably in recent years. Progress in quasicrystal growth~\cite{sc54},
the fabrication of quasiperiodic electronic structures such as Pb monolayers on quasicrystalline substrates~\cite{sc55}, the atom-by-atom
construction of synthetic electronic quasicrystals based on Penrose tilings~\cite{sc56}, the observation of superconductivity in the 
icosahedral Al-Zn-Mg quasicrystal~\cite{sc42}, and the recent realization of a superconducting moir\'e quasicrystal~\cite{sc57} suggest 
that superconducting hybrid devices incorporating quasiperiodic materials are within experimental reach. These developments provide a 
promising platform for directly testing the proximity-induced superconducting correlations predicted in the present work and, more 
generally, for exploring superconducting proximity effects in quasiperiodic systems.

\section*{ACKNOWLEDGMENTS}

S. K. acknowledges support from the University Grants Commission (UGC), India, through a research fellowship (NTA Reference No. 211610108524).
The authors sincerely thank Prof. Shreekantha Sil for valuable comments and insightful discussions.

\section*{DATA AVAILABILITY STATEMENT}

The data supporting the findings of this study are included in the manuscript. 


\section*{DECLARATION}

{\bf Conflict of interest} The authors declare no conflict of interest.

\end{document}